\documentclass[10pt,conference]{IEEEtran} 
\IEEEoverridecommandlockouts
\usepackage{amsmath,amssymb,amsfonts}
\usepackage{algorithmic}
\usepackage{graphicx}
\usepackage{textcomp}
\usepackage{xcolor}
\usepackage{url} 

\begin{document}

\title{Synthetic Test Data Generation Using Recurrent Neural Networks: A Position Paper}

\author{
	\IEEEauthorblockN{Razieh Behjati}
	\IEEEauthorblockA{\textit{Testify AS}, Oslo, Norway \\
		razieh.behjati@testify.no}
	\and
	\IEEEauthorblockN{Erik Arisholm}
	\IEEEauthorblockA{\textit{Testify AS}, Oslo, Norway \\
		erik.arisholm@testify.no}
\\
	\IEEEauthorblockN{Chao Tan}
	\IEEEauthorblockA{\textit{Dept. of Informatics, University of Oslo} \\
	\textit{Testify AS}, 
	Oslo, Norway \\ chao.tan@testify.no}
	\and
	\IEEEauthorblockN{Margrethe M. Bedregal}
	\IEEEauthorblockA{\textit{Skatteetaten}, Oslo, Norway \\
		margrethe.bedregal@skatteetaten.no}

	}
	\date{}

\maketitle

\begin{abstract}
Testing in production-like test environments is an essential part of quality assurance processes in many industries. 
Provisioning of such test environments, for information-intensive services, 
involves setting up databases that are rich-enough to enable simulating a wide variety of user scenarios. 
While production data is perhaps the gold-standard here, many organizations, particularly within the public sectors, are not allowed to use
production data for testing purposes due to privacy concerns. 
The alternatives are to use anonymized data, or synthetically generated data.
In this paper, we elaborate on these alternatives and compare them in an industrial context. 
Further we focus on synthetic data generation and investigate the use of recurrent neural networks for this purpose. 
In our preliminary experiments, we were able to generate representative and highly accurate data using a recurrent neural network. 
These results open new research questions that we discuss here, and plan to investigate in our future research.
\end{abstract}


\section{Introduction}
Governments, as well as many industries and organization require information-systems to be tested, end-to-end, in production-like environments before their release. These tests are in addition to other types of tests, such as unit tests, in the lower levels of the Test Automation Pyramid, introduced in~\cite{Cohn:2009}. 
Running these end-to-end tests requires building and configuring test environments that are similar to the production environment, 
with respect to the deployed components and the data available in their databases. 
Setting up such environments in the presence of dependencies to external data-providers is challenging. 
Therefore, to facilitate integration testing, many providers of information-intensive services offer test environments or sandboxes to their users.  
An example is the Norwegian National Registry, which releases electronic personal information to various legal entities in the country.

Currently, the Norwegian National Registry is undergoing a modernization process, and so are the consumers of its data. 
A board of quality assurance professionals, each representing one of the stakeholders, has decided to build a test environment dedicated to integration testing across these organizations and their systems. 
To the heart of this test environment is a single instance of the National Registry that, as in reality, disseminates \emph{the same data} to all consumers.
The choice of building a shared test environment may seem arbitrary,
but it is in fact essential for simulating realistic test scenarios and guaranteeing their correct execution. 
Suppose that the tax administration wants to execute a test scenario to verify the calculation of annual income taxes. 
This calculation requires fetching relevant information from other organizations, for instance the welfare system.
If the welfare system used for testing does not have access to the same National Registry data, it would most likely fail to respond correctly
to requests from the tax calculation system under test. 

Through a semi-structured interview~\cite{Simula-TR} with the main consumers of the Norwegian National Registry data, we have identified the main test-data needs that must be satisfied in the shared test environment described above:
\begin{itemize}
\item{\emph{Artificial data:}} To conform to privacy regulations, production data cannot be used for testing purposes. 
In addition, the level of anonymity provided by anonymized data available to the project is not sufficient for making it available to the external partners. 
The remaining alternative is synthetic data generation, which is considered as the most favorable option in this project.
\item{\emph{Dynamic data:}} National Registry data concerns people's personal information, and changes to them that happen due to the occurrences of life-events, such as birth, marriage, and death. Simulation of these events is essential for triggering the operations of consumer systems, which are the subjects of integration testing in this environment. 
\item{\emph{Representativeness:}} End-to-end tests are expensive and are not worth it if they cannot simulate realistic scenarios. 
The interaction between the dynamic nature of the data and the need for realistic test scenarios demands representativeness. 
Since the data in the test environment is constantly changing, hard-coding values in test scenarios is not advisable. 
To illustrate the problem, consider a test scenario for tax calculation, with a hard-coded personal identification number. 
This scenario may not produce the same results when executed several times, 
because the person might have left the country in between the executions (due to a synthetic emigration event), 
therefore being subject to different taxation rules. 
Instead, test scenarios must dynamically retrieve the test data suitable to their intent, using a search engine or similar mechanisms for querying the National Registry data. 
Such search engines may fail to find results matching all queries if the dataset is unrepresentative,
containing a relatively small number of combinations of personal data-attributes.
\end{itemize}

To address these needs, we are developing a synthetic data generator, which simulates occurrences of life-events. 
To ensure representativeness, we sample the life-events from probability distributions that summarize the statistical properties of the real Norwegian population.
These distributions are incorporated into a statistical model, which can be generated using different techniques. 
A promising approach, which we have briefly experimented with, is to use recurrent neural networks~\cite{Jain2000}. 
These are deep learning techniques, widely used in natural language processing applications, 
many of which involve synthesizing textual data.
So far, our data generator uses the statistical models only to generate what we call \emph{meta-events}, i.e., metadata about the events. 
A large code-base of glue logic is developed to process the meta-events and generate valid and consistent
fully-specified events in a format that can be written into the database. 
The impressive results of our preliminary experiment makes us wonder if we could retire the glue logic 
and replace the entire data generator with a single end-to-end deep learning model.    
This idea is analogous to the evolution of machine translation approaches, 
from statistical machine translation~\cite{SMT_03,SMT_90} that are complicated systems composed of many moving parts,
to neural machine translation~\cite{Google_NMT} that learn directly, in an end-to-end fashion, 
the mapping from input text to associated output text.

Other approaches for synthetic data generation are briefly discussed in the next section, 
along with a comparison to the use of anonymization techniques in our practical context. 
Section~\ref{sec:experiment} presents the results of our preliminary experiment. 
In the same section, we discuss the new research questions raised by the findings of our experiment.
Directions for future research indicated by these questions, plus a few concluding remarks are presented in the final section.

\section{Related work}

In the past, several researchers have focused on synthesizing statistically representative data.
For example, Soltana~\textit{et al.}~\cite{Soltana2018} model probabilistic characteristics of a population to generate synthetic data for simulating tax policies in Luxembourg. 
Synthesizing relational databases has been the focus of Chulyadyo and Leray~\cite{ChulyadyoL18} and Patki~\textit{et al.}~\cite{Patki2016}.
In their work~\cite{ChulyadyoL18}, Chulyadyo and Leray use probabilistic relational models to generate synthetic spatial datasets. 
Patki~\textit{et al.}~\cite{Patki2016} on the other hand, propose a general end-to-end synthetic data generator, named \emph{Synthetic Data Vault}, 
to synthesize complete tables of a relational database. The resulting databases resemble the original ones both statistically and structurally.
These earlier studies are not designed to generate dynamic data, 
and some of them, such as~\cite{Soltana2018}, require considerable manual effort to build the statistical models.  
Consequently, the applicability of these approaches is limited in our practical context, 
where the data generator is required to continuously generate new life-events that are expressed in terms of tens of interacting parameters.

In parallel to the aforementioned studies, deep learning techniques have been used in other domains for generating synthetic data.
The most commonly used techniques among these are recurrent neural networks~\cite{Jain2000} and generative adversarial networks~\cite{Goodfellow2014}. 
They have been used for synthesizing text~\cite{Sutskever2011,Rajeswar2017}, images~\cite{Gregor2015,Nie2017}, 
and sequential data~\cite{Graves2013,Ghosh2017}.
These approaches have no limitations with respect to the generation of dynamic data, 
and require no manual effort for generating the statistical models. 
However, they are computationally more expensive compared to the studies mentioned above.

When \emph{de-identification} is the goal, an alternative to synthetic data generation is anonymization. 
The benefit of anonymization is that it is suitable for generating dynamic data.
Having an anonymization algorithm, it is usually straightforward to create a streaming pipeline that continuously anonymizes new production data.
In addition, compared to synthetic data, anonymized data may better resemble the real data.
While anonymization algorithms hide the identities of real people to a great extent, 
they are not entirely immune to the so-called linking attacks, 
where certain attributes of individuals, such as age, gender, and zip code are joined to reveal a person's identity~\cite{Anonym05}.
Another drawback of using anonymized data is that it is not easy to downscale or upscale the dataset, 
as normally, anonymization algorithms map the real dataset to an anonymized dataset of the same size. 
This causes limitations for certain types of tests. 
For instance, performance tests may require larger amounts of data, 
while, smaller yet representative datasets are preferable for functional integration testing.  
Using synthetic data gives more flexibility with respect to these concerns.

\section{Synthetic Data Generation with Recurrent Neural Networks}
\label{sec:experiment}
Recurrent Neural Networks (RNNs)~\cite{Jain2000} are networks with loops in them, allowing information to persist.
This architecture equips RNNs with a form of internal state (memory) 
enabling them to exhibit temporal dynamic behaviors and to process sequences of inputs.
RNNs are widely used for natural language processing tasks, 
including machine translation~\cite{Google_NMT}, text synthesis~\cite{Sutskever2011}, e.g., as in chat-bots~\cite{Dialogue_resp,DeepProbe}, and speech recognition~\cite{speech_recognition_15}.
Inspired by these studies, we decided to design an experiment to evaluate the effectiveness of RNNs in synthesizing meta-events. 

In our experiment, we use a Long Short-Term Memory (LSTM)\cite{LSM-1997}, a special RNN architecture, to build a \emph{generative model}. 
In general, a generative model captures the joint probability, $p( x, y)$, of the inputs $x$ and the output $y$. 
This probability can then be used to sample data, or to make predictions by using Bayes rules to calculate the posterior probability $p(y | x)$, 
and then picking the most likely output $y$.
In contrast, \emph{discriminative models} capture the posterior probability $p(y | x)$ directly, 
and are used for making predictions in classification and regression tasks\cite{Ng:2001}. 

In the rest of this section, we first explain our experimental setup, and the dataset we used for training. 
The results of our experiments are presented in Section~\ref{ssec:res}, followed by a discussion of the findings in Section~\ref{ssec:discussion}. 

\subsection{Experiment}
For this experiment, we used the code written by Karpathy, available at~\cite{char_rnn}.
The model takes one text file as input and trains an LSTM that learns to predict the next character in a sequence. 
The model is then used to generate text, character by character, that will look like the original input data.

The dataset for this experiment is about 42 Megabytes, and contains 141 570 fixed-length records.
Each record is 300 characters long and represents a meta-event,
which includes the birthdate of the person involved in the event, the date and time of the event, 
a code representing the type of the event, 
and several other codes each representing a particular detail about the person (e.g., marital status, and residence status). 
Other personal details, such as family relations, addresses, and last updated dates of each of these fields 
will be encoded as part of the meta-events in upcoming releases of the data generator. 

We used 0.6/0.1/0.3 ratios to split the data into training, validation, and test, respectively\footnote{In classification and regression tasks the test dataset is used to provide an unbiased evaluation of a final model fit on the training dataset and tuned based on the validation dataset. 
The performance of the model is measured based on how well it can predict the target values for the test dataset. 
In the experiment reported here, we are facing a data generation task as opposed to a prediction task, so the test dataset has no use here.}. 
The total number of distinct characters in the dataset is 53. 
This dataset is collected from a test environment used internally in the Modernization project of the Norwegian National Registry.
The dataset itself is, therefore, synthetic, generated manually or through the execution of automated test suites, in the course of several years.
As a result, the statistical characteristics of this dataset are very different from those of the real Norwegian National Registry data.

The network we used in this experiment is a 2-layer LSTM with 256 units in each layer. 
The network takes the data as batches of size 20. 
Each data point in the batch is a sequence of 300 timesteps, and there is a one-hot layer before feeding the data to the LSTM cells. 
Therefore, the input tensors to the LSTM layer have the shape $(20, 300, 53)$.
The expected output is also a sequence of length 300, generated from the same original text, but shifted one character to the right. 
These tensors are converted to one-hot representations before being consumed by the network. 

The network has a total of 858 421 trainable parameters, and was trained on an Ubuntu machine with one NVIDIA Tesla M60 GPU. 
The training terminated after completing 16 iterations over the entire training set, taking about 12 hours. 

The trained model was then used to sample $6.3$ Megabytes of data, containing 20 844 records.
At each step during sampling, the network generates a distribution over what characters are likely to come next. 
This distribution is then used to sample a new character, which is fed back into the network to get the next character. 
This process is repeated until the desired number of characters are generated.

\subsection{Results\label{ssec:res}}
The sampled meta-events will be used as input to our data generator. 
Therefore, it is important for them to be well-formed with respect to the format assumed by our data generator.
In addition, we want the distribution of the sampled data to be similar to that of the original dataset.
The following analysis provides an insight into how the sampled dataset addresses these two requirements. 

\subsubsection{Well-formedness}\hfill

\begin{figure}
    \centering
    \includegraphics[width=0.8\linewidth]{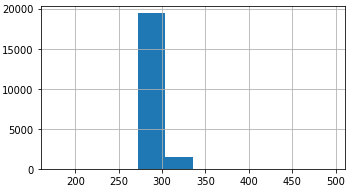}
    \caption{Distribution of the sampled records lengths}
    \label{fig:length}
\end{figure}

In the first step of our analysis we examined the length of the sampled records. 
As shown in Figure~\ref{fig:length}, almost all the sampled records have a length close to 300. 
In fact, the lengths of 99.74\% of the sampled records lie within the $300\pm15$ range. 

Meta-events used in this experiment start with 11-digit personal identification numbers
that encode the birthdate of the person, and end in 2-digit Modulo-11 check-sums.
We examined the validity of the sampled identification numbers according to this rule. 
Among the 20~772 unique ids, only 169 of them are valid identification numbers;
but 20~497 ($\sim98.8\%$) of the ids are invalid only because the check-sum digits do not match the rest of the string. 
In other words, more than $98.8\%$ of the sampled identification numbers encode a valid date as the birth date.
Clearly, the Modulo-11 algorithm is too complex to be learned by this model.

The characters in positions 11-25 of each meta-event indicate the time-stamp of the event, 
and should be able to be parsed into a date-time object. 
In the sampled dataset, 99.7\% of the records have a correctly formatted time-stamp. 
The other 51 records, either have a day that is out of range for month (e.g., February 29 in a nonleap year), 
or an hour or a minute that is greater than 60. 
A simple policy to deal with these cases is to ignore them. 
Since these cases form less than $0.3\%$ of the samples, 
and the validity check is not expensive such a policy would not have a huge impact on the performance of the event generator.

\subsubsection{Representativeness}\hfill

To verify the representativeness of the model, 
we compare the sampled dataset with our original dataset\footnote{Note that here we have not used the original dataset as input for creating the sampled dataset. 
The sampled dataset is generated one character at a time, starting from an empty string. 
Therefore, the comparison to the original dataset does not result in a biased evaluation of the performance.}
both visually, and quantitatively 
using the Jensen-Shannon divergence method~\cite{Lin1991Divergence},
which is a method for measuring the similarity between two probability distributions.
This metric is equal to zero only if the two probability distributions are identical, 
and grows to an upper bound, which in our implementation is $ln(2)$, as the two distributions diverge.

\begin{figure*}
    \centering
    \includegraphics[width=\linewidth]{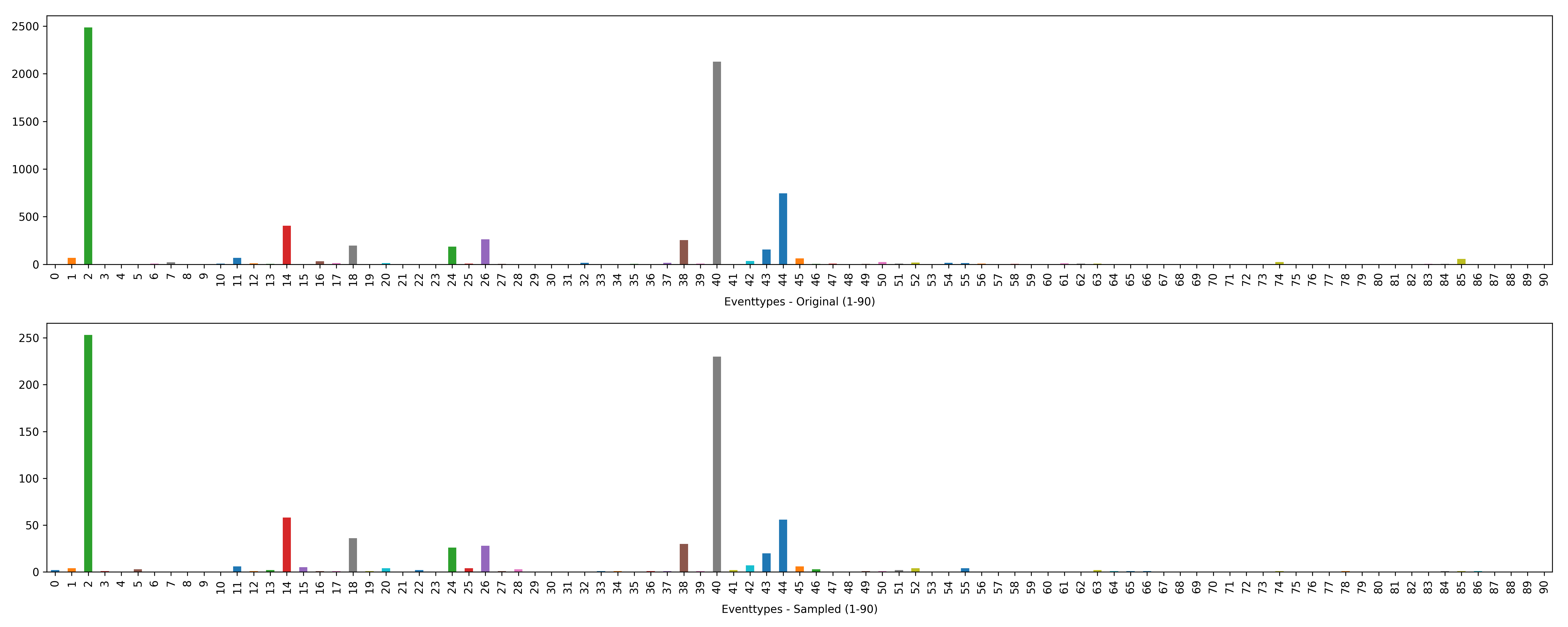}
    \caption{Distribution of the event types. 
    For the sake of better visualization, events coded as 91 and 92 are excluded from this diagram. 
    These two codes have very high counts in both original ($\thicksim$134 000) and sampled ($\thicksim$19 000) datasets. 
    They are, however, included in the calculation of Jensen-Shannon divergence reported in the text ($0.002 087$).
    Excluding these event codes results in a Jensen-Shannon divergence of  $0.028 685$, which is still quite low, 
    but much higher than the original value, indicating less similarity between the two distributions in the 1-90 range. 
    This demonstrates the impact of the data imbalance on the capability of the model in generating representative data for the less frequent classes.
    This degree of imbalance however is not present in the real Norwegian National Registry data. 
    It is present in the dataset reported here because several load tests, 
    generating thousands of code 91 and 92 events, were executed in the test environments from which we collected the data for this experiment.
    }
    \label{fig:eventtypes}
    \vspace{1cm}
\end{figure*}

Figure~\ref{fig:eventtypes} compares the distributions of the event types in the original (top) and sampled (bottom) datasets. 
The labels on the horizontal axis are the codes that represent the event types. For instance, $1$ is the code for a birth event.
The Jensen-Shannon divergence for these two distributions is equal to $0.002 087$.
These results indicate that the network has manged to preserve the distribution of the event types to a great extent. 
There are certain types of events that did not exist in the original dataset, but were generated by the model (e.g., 0, 5, 41); 
and types of events that existed in the original dataset but were not sampled by the model (e.g., 7, 23, 47). However, all these are very infrequent.
This observation suggests that the network is incapable of distinguishing between very low probabilities, 
and a probability of zero that may indicate that a value or a combination of values are invalid.

\begin{figure*}
    \centering
    \includegraphics[width=\linewidth]{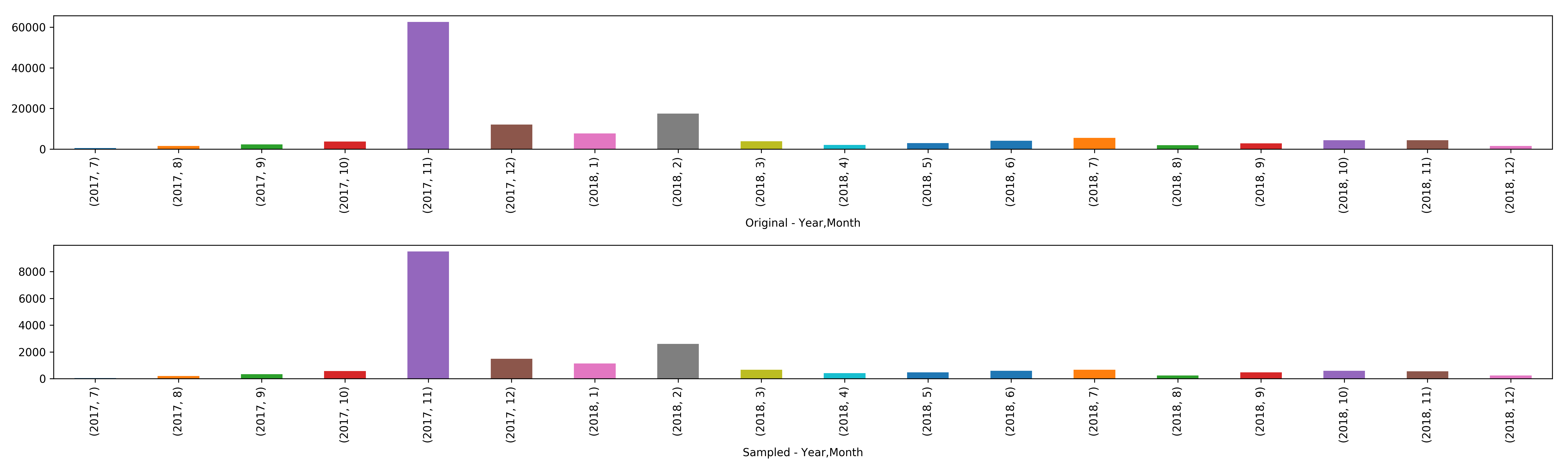}
    \caption{Distribution of the time-stamps}
    \label{fig:timestamps}
    \vspace{0.5cm}
\end{figure*}

Finally, Figure~\ref{fig:timestamps} compares the distributions of the time-stamps in the original and sampled datasets. 
The Jensen-Shannon divergence for these two distributions is equal to $0.001 142$.
These results suggest that meta-events in the sampled dataset have a temporal distribution very similar to that in the original dataset.

\subsection{Discussion\label{ssec:discussion}}
The results presented above show that the recurrent neural network used in this experiment 
has been able to very accurately model the structure of the data. 
It knows the length of the records, the location of each field, and the format of simple constructs such as dates and times.
Further, it has learned simple dependencies and patterns, such as the valid ranges for month values, 
and correlations such as the upper bounds for the day of month. 
In addition, the accuracy of the modeled probability distributions is remarkable. 

However, the model has failed to learn more complex patterns, such as the Modulo-11 check-sum calculation, 
or the calculation of the leap years. The latter is likely to be due to insufficient training data.
Another drawback of this model is that it cannot simulate events that would happen after December 2018. 

As mentioned earlier, our data generator currently contains many lines of 
glue logic code required for converting the meta-events into fully-specified events, 
and verifying that they are valid, and correctly formatted, such that they can be written into our databases,
without causing any inconsistencies. 
The ultimate goal of our research is to reduce this glue logic as much as possible, 
and replace (parts of) it with machine learning solutions that are easier to maintain, and can generate meta-events 
that are both syntactically and semantically more similar to the final fully-specified events.
For this purpose, in addition to addressing the shortcomings mentioned above, 
we will augment the meta-events with more data fields, 
such as those concerning the relationships or earlier states of the person. 
Consequently, in our future research we will focus on the following research questions:
 \begin{itemize}
 \item \textbf{RQ1:} \emph{To what extent can deep neural networks learn complex patterns within and across events?}
Examples of patterns within an event are the calculations of Modulo-11 check-sum, and leap years noted above. 
For a pattern that spans across multiple events, consider a marriage life-event.
In the National Registry, each registered event concerns a single person. 
So, in the case of a marriage, two separate events need to be recorded in the database, one for each spouse. 
But these two events share a lot of information (such as the date and location of marriage), 
and contain references to each other; 
namely each event has a field for the identification number of the spouse that, in this case, is equal to the subject of the other event. 
There are other similar cases, where a group of two or more related events need to be generated together 
in order to preserve the integrity of the data. 
It would be interesting to see whether existing deep neural architectures reported in the literature, 
in particular variations of RRNs (e.g., Bilateral-LSTM~\cite{Bilat-LSTM}),
can learn the logic behind the grouping of events, 
or novel architectures need to be designed for this purpose.
 
 \item \textbf{RQ2:} \emph{To what extent can deep neural networks learn the rules governing the applicability of each event? }
Many events are only valid under specific situations. For example, relocation of a dead person is meaningless, and therefore is not allowed.
In state-transition models, these rules are referred to as transition constraints or guards. 
They express a relationship between the current state of a person, and the events that can happen in that state. 
A straightforward solution to allow learning these constrains is to encode the current state of the person as part of the meta-events used for training. 
In addition to this, we need to solve the problem of the incapability of the network to distinguish between very low probabilities, and the zero probability, addressed above.
Supervised learning, and reinforcement learning approaches are the commonly used solutions in situations like this.

This is perhaps our most important research question, 
since every type of event is involved in some state-transition constraints, 
and a sub-optimal approach for modeling the constraints can risk the validity of every single event generated by our tool.
Add to this the fact that there is very little tolerance for inconsistencies within the test data in the  test environment  offered by the Norwegian National Registry.
 
 \item \textbf{RQ3:} \emph{Can deep neural networks extrapolate future events?}
 As discussed above, the RNN used in this experiment is not able to simulate events that occur after December 2018.
 In fact, in our context, in addition to the temporal distribution of the events, their chronological order is also important. 
 To address this problem, we plan to investigate the use of conditional generative adversarial networks~\cite{CON_GAN}.
 These networks can control certain aspects of the sampled data, by feeding into the network input noise-vectors that are generated from a specific a-priori distribution. 
 \end{itemize}

\section{Conclusion}
In this paper we discussed the problem of synthetic test data generation for event-driven systems. 
We presented a set of requirements, including the need for valid, and representative dynamic data.
Results of a preliminary experiment using a Long Short-Term Memory network, reported in this paper, 
show that these deep neural architectures can provide promising solutions to address both of the requirements.
In the future, we will further evaluate the use of recurrent neural networks on a lager dataset with more complex patterns and dependencies.
We consider the use of supervised and reinforcement learning techniques for enforcing more complex validity constraints.
Furthermore, we will investigate the use of generative adversarial networks for sampling future events: 
events that happen on dates that are not present in the training dataset.

\section*{Acknowledgment}
We thank the Modernization project of the Norwegian National Registry, 
and the Norwegian Tax Administration that hosts the project and is responsible for keeping the national registry updated.

\nocite{*}
\bibliographystyle{abbrv}
\bibliography{refs}

\end{document}